\documentclass[twocolumn]{autart}

\usepackage[round]{natbib}
\usepackage{amsmath,bm}
\usepackage{color}
\usepackage{multirow}
\usepackage{mathrsfs}
\usepackage{tcolorbox}
\usepackage{subfig}
\usepackage{dsfont} 
\usepackage{amssymb}
\usepackage{setspace}
\usepackage{graphicx}
\usepackage{float}
\newtheorem{remark}{Remark}
\newtheorem{lemma}{Lemma}

\newtheorem{proposition}{Proposition}

\newtheorem{proof}{\textbf{Proof}}
\def\begcen{\begin{center}}
\def\endcen{\end{center}}
\newcommand{\col}{\mbox{col}}

\def\calp{{\mathfrak p}}

\def\L2e{{\mathcal L}_{2e}}

\def\rea{\mathbb{R}}

\def\l2{{\mathcal L}_2}
\def\l2e{{\cal L}_{2e}}

\def\rea{\mathbb{R}}

\def\begequarr{\begin{eqnarray}}
\def\endequarr{\end{eqnarray}}
\def\begequarrs{\begin{eqnarray*}}
\def\endequarrs{\end{eqnarray*}}
\def\begarr{\begin{array}}
\def\endarr{\end{array}}
\def\begequ{\begin{equation}}
\def\endequ{\end{equation}}
\def\lab{\label}
\def\begdes{\begin{description}}
\def\enddes{\end{description}}
\def\begenu{\begin{enumerate}}
\def\begite{\begin{itemize}}
\def\endite{\end{itemize}}
\def\endenu{\end{enumerate}}

\def\lef[{\begin{array}}
\def\rig]{\end{array}}
\def\qed{\hfill$\Box$}
\def\begcen{\begin{center}}
\def\endcen{\end{center}}
\def\begrem{\begin{remark}\rm}
\def\endrem{\end{remark}}
\def\begali#1{\begin{align}{#1}\end{align}}
\def\begalis#1{\begin{align*}{#1}\end{align*}}

\def\begsubequ{\begin{subequations}}
	\def\endsubequ{\end{subequations}}
\usepackage{xcolor}

\begin{document}
\begin{frontmatter}

\title{Standard LS Parameter Estimators Ensure Finite Convergence Time for Linear Regression Equations Under an Interval Excitation Assumption} 

\author[ITAM]{Romeo Ortega},
\author[ITAM]{Jose Guadalupe Romero},
\author[SUP]{Stanislav Aranovskiy},
\author[VIR]{Gang Tao}

\address[ITAM]{Department of Electrical and Electronic Engineering, ITAM, Progreso Tizap\'an 1, Ciudad de M\'exico, 01080, M\'{e}xico}
\address[SUP]{IETR Laboratory, Centrale Sup\'elec, 35510
Cesson-S\'evign\'e, France }
\address[VIR]{Department of Electrical and Computer Engineering, University of Virginia, Charlottesville, VA 22903, USA}

\begin{abstract}
In this brief note we recall the little-known fact that, for linear regression equations with intervally excited (IE) regressors, {\em standard} Least Squares parameter estimators ensure {\em finite convergence time} of the estimated parameters. The convergence time being equal to the time length needed to comply with the IE assumption. As is well-known, IE is necessary and sufficient for the {\em identifiability} of the linear regression equation---hence, it is the weakest assumption for the on- or off-line solution of the parameter estimation problem. 
\end{abstract}


\end{frontmatter}
%
%%%%%%%%%
\section{Introduction}
\lab{sec1}
%%%%%%%%%%5
%
In the last few years we have seen a significant growth of interest in the development of algorithms (for optimization, stabilization, state observation or parameter estimation) with {\em finite (or fixed) convergence time} (FCT). With some notable exceptions, the vast majority of these algorithms appeal to signal differentiation and/or injection of {\em high-gain} (HG) in the control loop---hence, this research line has mainly attracted control {\em theoreticians}. 

Researchers interested in practical applications are well aware of the disastrous amplification effect of differentiation and/or HG  on the inevitable presence of noise---see, {\em e.g.}, \citep{ARAetal} for a recent example of this unavoidable phenomenon. Moreover, the claim of FCT is, in reality, specious in applications since---due to the presence of noise---the best we can expect is that the controlled trajectory enters a band determined by the noise amplitude. It is pertinent to note that the claim of ``robustness to external disturbances" of the prototypical HG  injection scheme of {\em sliding mode} (SM) control (that injects in the loop HG  via the incorporation of a {\em relay}, which is an infinite gain operator) refers only to the case of ``matched disturbances", {\em i.e.}, those entering into the image of the input matrix disregarding the output measurement noise---a clarification often omitted in the SM literature. 

Algorithms that achieve FCT have been reported in almost all control problems, with the FCT objective achieved via differentiation and/or the injection of HG. An early summary of the results on this topic may be found in \citep{BAS}, where schemes with SMs, supertwisting and differentiation operators, are discussed. Another reference, that concentrates on the role played by {\em homogeneity}---introduced to the control community by \citep{BHABER,KAW}---is given in \citep{EFIPOL}. It is interesting to note that the kind of nonlinearities that appear in most physical systems are {\em transcendental} functions, {\em e.g.}, sinusoidal, exponential, logarithmic, and {\em are not homogeneous}---hence, the results reported in  \citep{EFIPOL} are of limited interest for them. HG  injection is also used in classical optimization problems to achieve FCT. For instance, in \citep{CHEWANWAN} a search gain with a normalized gradient algorithm is defined via a Caputo fractional derivative. The highly fashionable topic of consensus control has also been addressed with HG  schemes to achieve FCT. For instance, in \citep{HUAetal} SM-like signum functions are added in the control loop. The use of HG  injection via the addition of {\em fractional powers} has also had a wide spread popularity in several control problems, including observer design \citep{ANDPRAAST}, control of nonlinear systems \citep{CRUNUNMOR, MINetal} and parameter estimation \citep{RIOetal,WANEFIBOB}---it is interesting to note the deleterious effect of noise in these estimators, vividly shown in \citep[Figs. 2 and 4]{WANEFIBOB}. 

As mentioned above, the research on FCT control schemes has, mainly, been driven by (mathemathically skillful) control theoreticians, whose involved derivations usually lead to highly complex algorithms. As an illustration of this point the reader is referred to the FCT state observer for a {\em linear time-invariant (LTI) observable} $(A,B,C)$-system proposed in \citep[Theorem 5.1]{EFIPOL}, namely:
$$
\dot {\hat x}  =A \hat x+Bu+g(C\hat x - y),
$$
where $g: \rea^k \to \rea^n$ is a map defined as
$$
g(s)=-{\gamma \over 2}|\gamma s|^\nu \exp(\ln |\gamma s|^\nu G_0)P^{-1}C^\top s
$$
with $P \in \rea^{n \times n}$ and $G_0 \in \rea^{n \times n}$ matrices satisfying an algebraic \citep[equation (5.4)]{EFIPOL} and a linear matrix inequality \citep[equations (5.5), (5.6)]{EFIPOL}, $\gamma>0$ and $\nu <0$. This observer, compared with the classical Luenberger observer $g(s)=-Ls$, can hardly be classified as ``easily implementable". See also \citep[Theorem 4.4]{EFIPOL} for a similarly complex {\em state feedback} controller for FCT stabilization of a {\em controllable LTI} system. 

Our interest in this paper is on the topic of on-line, continuous-time, {\em parameter estimators} with FCT, where the unknown parameters satisfy a {\em linear regression equation} (LRE) of the form
\begequ
\lab{lre}
y(t)=\phi^\top(t) \theta,
\endequ
where $y(t) \in \rea$ and $\phi(t) \in \rea^{q}$ are {\em measurable} signals and $\theta \in \rea^q$ is a vector of {\em unknown, constant} parameters.\footnote{We consider, without loss of generality, the simplest case of scalar signal $y(t)$---the extension to the vector case follows {\em verbatim}.} In particular, we are interested in {\em least-squares} (LS) algorithms \citep{RAOTOUbook}---the interested reader is referred to \citep[Subsection 1.1]{ORTROMARA} for a review of the recent results in LS parameter estimators. 

The rest of the paper has the following structure. Some historical preliminaries on LS estimators are given in Section \ref{sec2}. The main result, the proof that standard LS with IE regressors yields an FCT algorithm, is given in Section \ref{sec3}. In Section \ref{sec4} we compare in simulations the performance of the following estimators: FCT-LS, standard LS and the FCT estimator with HG  injection reported in \citep{WANEFIBOB}. 

%
%%%%%%%%%%%
\section{Historical Preliminaries}
\lab{sec2}
%%%%%%%%%%%%%
%
To the best of our knowledge the first time that a claim of FCT for LS estimators was published in \citep{ORTproieee}---reported again 20 years later in \citep{ADEGUA}. The algorithm in \citep{ORTproieee} was motivated by the normalization results of \citep{KRAKHA} and has several major numerical drawbacks, thoroughly discussed in \citep{PANSHIORT}. 

A major breakthrough in the understanding of LS estimators is the observation that $F^{-1}(t) \tilde \theta(t)$ is {\em constant}, with $F(t) \in \rea^{q \times q}$ the covariance matrix and $\tilde \theta(t) \in \rea^q$ the parameter error. This observation was made in \citep[Equation 17]{DEL} for the case of discrete-time LS estimators without forgetting factor, and it was extensively exploited for the design of indirect adaptive controllers \citep{LOZZHA}. To illustrate such an observation, we use an unnormalized LS algorithm without a forgetting factor:
\begalis{
\dot{\hat{\theta}}(t) & =  F(t) \phi(t) [y(t) - \phi^\top(t) \hat{\theta}(t)] \\
\dot{F}(t) & =  -F(t) \phi(t) \phi^{\top}(t) F(t),\;F(0) = F^\top(0) > 0.
}
Computing now
\begalis{
{d \over dt } &\Big(F^{-1}(t) \tilde{\theta}(t)\Big)  = F^{-1}(t)\dot{ \tilde{\theta}}(t)+ \dot F^{-1}(t) \tilde{\theta}(t) \\
& \quad =- F^{-1}(t) F(t)\phi(t) \phi^{\top}(t) \tilde{\theta}(t)+\phi(t) \phi^{\top}(t) \tilde{\theta}(t) \\
 & \quad =  0,
}
where we have used \eqref{lre} in the first right hand term of the second equation and the fact that $\dot F^{-1}(t)=\phi(t) \phi^{\top}(t)$ in the second term.

An immediate consequence of this fact is that
\begequ
\lab{keyequ}
\tilde \theta(t)=F(t) F^{-1}(0)\tilde \theta(0),
\endequ
and equation that appears in \citep[equation (17)]{DEL} and later on \citep[equation (8.108)]{SLOLIbook} for an (unnormalized) LS {\em with forgetting factor}. 

Two key steps, carried out in \citep[Lemma 3.5]{TAObook}, that lead to the FCT result are:
\begenu
\item[(i)] the rearrangement of \eqref{keyequ} in the form
$$
[I_q-F(t) F^{-1}(0)]\theta=\hat \theta(t)-F(t) F^{-1}(0) \hat \theta(0);
$$
\item[(ii)] the observation that the following implication is true 
\begin{align}
\lab{ie}
\exists T_c>0,\;\rho>0: & \int_0^{T_c} \phi(\tau)\phi^\top(\tau)d\tau \geq \rho I_q\\\
\nonumber \Rightarrow \; & 
\det\{I_q-F(t) F^{-1}(0)\} \neq 0,\;\forall t \geq T_c.
\end{align}
\endenu

A regressor $\phi(t)$ satisfying the inequality above is said to be IE,\footnote{Called ``exciting over an interval" in \citep[Definition 3.1]{TAObook}, where the limits of integration are taken as $[\sigma_0,\sigma_0+T_c]$, with $\sigma_0\geq 0$.} a term first coined in  \citep{KRERIE}. The same steps were carried out for  (continuous- and discrete-time) LS with forgetting factor in \citep{ORTROMARA}, but the connection with FCT was not established.
%
%%%
\section{Standard LS Estimation has FCT}
\label{sec3}
%%%%%%%%%%%%
%
In this section we present our main result, namely, the proof that standard LS yields an FCT estimator.

\begin{proposition}
\lab{pro1}\em
Consider the regression equation \eqref{lre} and assume $\phi(t)$  is IE and {\it bounded}.  Define the standard LS estimator with forgetting factor \citep[Subsection 8.7.6]{SLOLIbook}.
\begsubequ
\lab{lsd}
\begali{
\lab{lsd1}
		\dot{\hat \theta}(t) & =\gamma_F F(t) \phi(t)   [y(t)-\phi^\top(t) \hat\theta(t)]\\
\lab{lsd2}
\dot {F}(t)& =  -\gamma_F F(t) \phi(t) \phi^\top(t)   F(t) +\chi(t) F(t)\\
\lab{lsd3}
\dot z(t) &=\; -\chi(t) z(t)\\
\chi(t) &= \chi_0 \left( 1-{{\| F(t)\|}\over{k}} \right)
}
\endsubequ
with initial conditions $\hat\theta (0)=\theta_{0} \in \rea^{q},\; F(0)={1 \over f_0} I_{q}$, $z(0)=1$ and tuning gains the scalars $\gamma_F>0$, $f_0>0$, $\chi_0> 0$ and $k\geq \frac{1}{f_0}$. 

For $t \geq T_c$, define the signal
\begequ
\lab{thefct}
\theta _{FCT}(t):=[I_q-z(t)f_0F(t)]^{-1}[\hat \theta (t)-z(t)f_0F(t) \theta _0].
\endequ
\begenu
\item [{\bf (i)}] 
For all initial conditions this signal verifies
$$
\theta _{FCT}(t)= \theta ,\;\forall t \geq T_c.
$$
\item [{\bf (ii)}] All the signals are {\it bounded}.
\endenu 
\end{proposition}

\begin{proof} \em
The result is contained in \citep[Lemma 3.5]{TAObook} for (normalized) LS {\em without} forgetting factor and, for LS+Dynamic Regression Extension and Mixing (DREM) of \citep{ORTROMARA} with forgetting factor, it follows directly from the proof of \citep[Proposition 1]{ORTROMARA}, where it is shown that the matrix $I_q-z(t)f_0F(t)$ is full rank for $t \geq T_c$.
\qed
\end{proof}

Regarding the IE assumption it is important to recall the lemma below \citep{WANetal}, which shows that the IE assumption is {\em necessary and sufficient} to estimate the parameter $\theta$ from the LRE \eqref{lre} with {\em on-  or off-line} estimators. 

\begin{lemma}\em
\lab{lem1}
The LRE \eqref{lre} is {\em identifiable\footnote{See \citep[Definition 2.2]{WANetal} for the definition of ``identifiable" LRE.}} {\em if and only if} the regressor vector $\phi(t)$ is IE.
\end{lemma}
%
%%%%%%%%%%%5
\section{Simulation Results}
\lab{sec4}
%%%%%%%%%%%%%%%5
%
 In this section we present simulations of the proposed FCT-LS  estimator and compare its performance with the standard asymptotically convergent LS one and with the HG  based one reported in \citep{WANEFIBOB}---in particular, in the face of measurement noise.
 \subsection{Example 5 of \citep{MARTOM}}
%%%%%%%%%%%5
We consider here Example 5 from \citep{MARTOM} to show that the new proposal has a better behavior in comparison to standard LS and HG -based techniques.

Consider the second order stable linear system described by 
\begin{align*}
\dot x_1(t)=&x_2(t) \\
\dot x_2(t)=& -\theta_1 x_1(t) -\theta_2{x_2(t)} +\theta_3 u(t) %\\ y(t)=&x_1(t),
\end{align*}
or equivalently 
\begin{equation} 
\ddot x_1(t) = -\theta_1  x_1(t)-\theta_2 \dot x_1(t)  +\theta_3 u(t)
\label{ex5MT}
\end{equation}
where $\theta_1$, $\theta_2$ and $\theta_3$ are unknown parameters. Applying to both sides of \eqref{ex5MT} the filter 
$$
H(\calp)={1 \over {\calp+\lambda}}
$$
where $\calp:=\frac{d}{dt}$ and $\lambda>0$,  and rearranging the terms, we get the LRE \eqref{lre} with 
\begin{align}
y(t) & =  \calp H({\calp}) [x_2(t)] \nonumber \\
\phi(t) &=H({\calp})[\col\left( -x_1(t), \, -\calp x_1(t), \, u(t) \right)] \nonumber % \label{eq:phi}
\end{align}
and $\theta:=\col(\theta_1, \, \theta_2, \, \theta_3)$.

{In what follows, we compare the standard asymptotic LS estimator \eqref{lsd}, the FCT estimator \eqref{thefct} and the HG-based estimator given by (9) of \citep{WANEFIBOB}.

For the system \eqref{ex5MT}, we use the same conditions of \citep{ORTROMARA}, that is, we choose $\lambda=1$ and set to zero the initial conditions of the filters. The vector of unknown parameters is chosen as $\theta=\col(2,3,1)$, and the input signal is set $u(t)=5$. Such input signal provides only interval (but not persistent) excitation to the regressor $\phi$, allowing for FCT parameter estimation.

To carry out the simulations of \eqref{lsd} and \eqref{thefct}, we chose the initial value of the parameter estimation vector as $\hat \theta(0)=\col(0.1,\,0.1, \, 0.1)$ and set the tuning parameters of the proposed estimator to $\gamma_F=30.3$, $f_0=4$, $\chi_0=6$ and $k=10$. The FCT estimate \eqref{thefct} is computed for time instances when $\det\left(I_q-z(t)f_0F(t)\right)\ge \delta_{FCT}$, where $\delta_{FCT}$ is set to $0.001$. For the HG-FCT estimator of \citep{WANEFIBOB}, we set $\beta_1 = 7$, $\beta_2 = 3$, $\alpha_i=10\beta_i$ for $i=1,2$, $\alpha = 0.5$, and we set the gain $\gamma$ and the initial conditions to the same values as for the FCT-LS estimator.}

\begin{figure}
	\centering
	\subfloat{\includegraphics[width=0.95\linewidth]{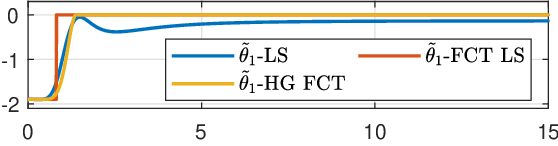}}
	\\
	\subfloat{\includegraphics[width=0.95\linewidth]{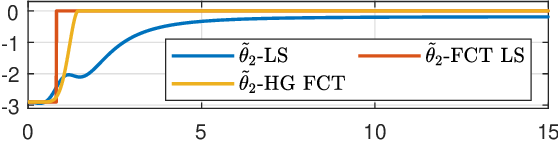}}
	\\
	\subfloat{\includegraphics[width=0.95\linewidth]{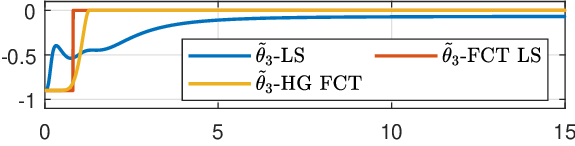}}
	\caption{Transient behavior of the estimated parameters $\tilde{\theta}_i(t)$ with $i=1,2,3$.}
	\label{fig1}
\end{figure}

In Fig. \ref{fig1} we appreciate the transient behavior of the estimation errors $\tilde{\theta}_i$ with $i=1,2,3$ using the standard LS \eqref{lsd}, the new FCT-LS \eqref{thefct}, and the HG-FCT estimator. As expected, the LS estimator does not converge under the IE condition, whereas both FCT-LS and HG-FCT exhibit FCT. 

\begin{figure}
\centering
  \includegraphics[width=.95\linewidth]{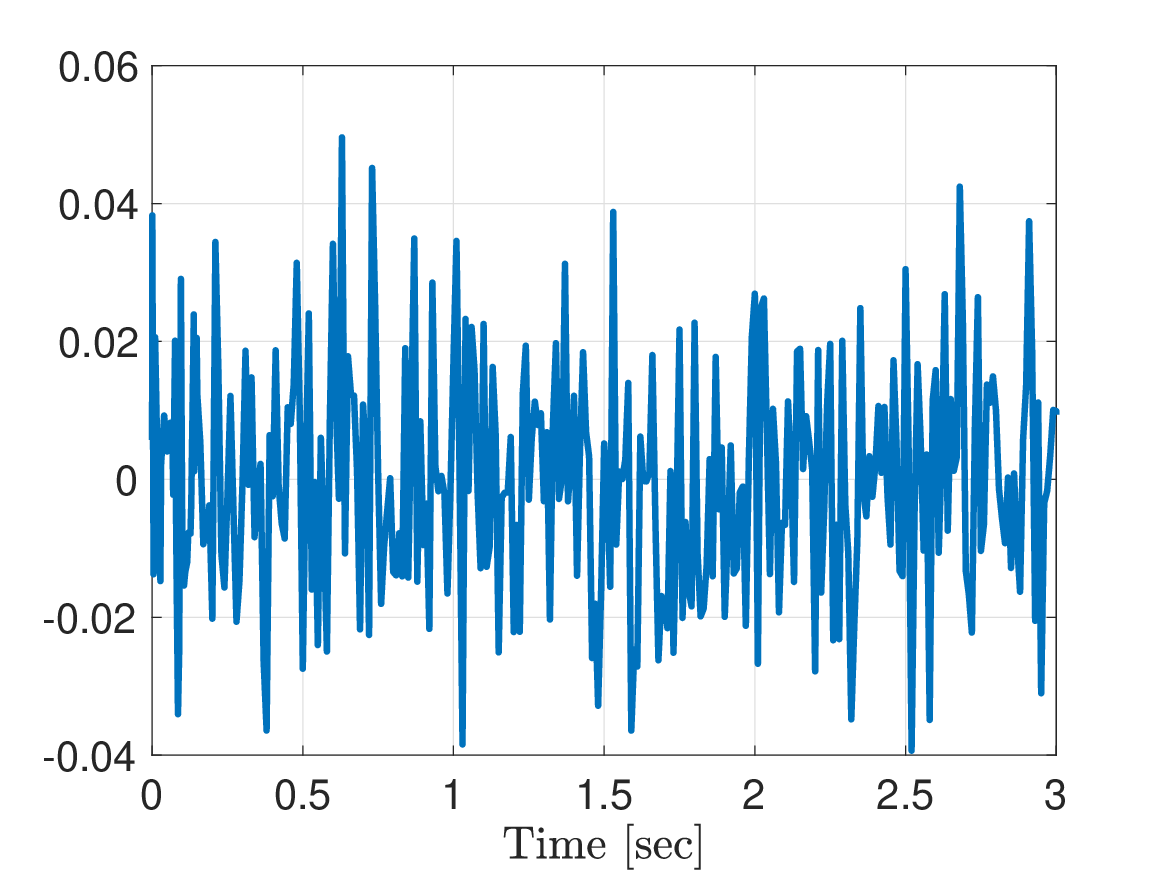}  
  \caption{ White noise signal}
 \label{fig2}
\end{figure}

\begin{figure}
	\centering
	\centering
	\subfloat{\includegraphics[width=0.95\linewidth]{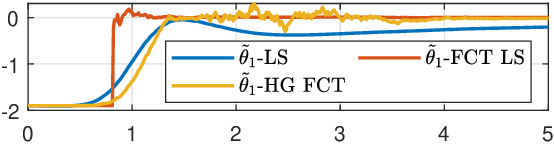}}
	\\
	\subfloat{\includegraphics[width=0.95\linewidth]{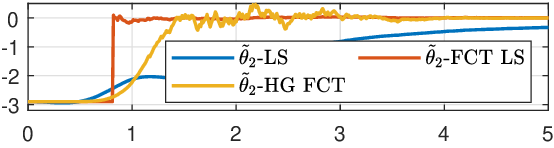}}
	\\
	\subfloat{\includegraphics[width=0.95\linewidth]{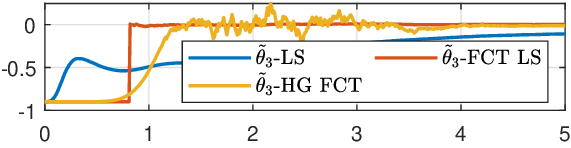}}
	\caption{Transient behavior of the estimated parameters  $\hat \theta_i(t)$ with $i=1,2,3$ with noise.} 
	\label{fig3}
\end{figure}

On the other hand, it is widely accepted that control and estimation methods that depend on HG injection are susceptible to noise. To illustrate the robustness of estimators, we repeat the simulations by adding a white noise signal (Fig. \ref{fig2}) of small amplitude to the measurable signal $y(t)$. The result of the simulation is shown in Fig. \ref{fig3}. We notice that even in the presence of noise, the estimation errors of FCT estimators oscillate around zero. However, the proposed FCT-LS estimator has significantly smaller magnitude of oscillations compared to the HG-FCT estimator. 

\bigskip
 
%\begin{figure}[htp]
%\centering
%  \includegraphics[width=.95\linewidth]{figures/fig3}  
%  \caption{Transient behavior of the estimated parameters  $\hat \theta_i(t)$ with $i=1,2,3$ with noise. \rom{Change LS-DREM to LS}} 
% \label{fig3}
%\end{figure}

%Finally, in Fig. \ref{fig4} we show the behavior of the HG-based estimator of \rom{Ref} in the presence of a measurement noise.

%\begin{figure}[htp]
%\centering
% % \includegraphics[width=.65\linewidth]{figures/fig4}  
%  \caption{Transient behavior of the estimated parameters  $\hat \theta_i(t)$ with $i=1,2,3$ with noise of the HG-based estimator of \rom{Ref}.}
% \label{fig4}
%\end{figure}
%
%%%%%%%%%%%%%%
%
%%%%%%%%%%%5
\section{Concluding Remarks}
\lab{sec5}
%%%%%%%%%%%%%%%5
%
There are many parameter estimators for LRE reported in the literature that ensure FCT without the injection of high-gains. A recent survey of some of them may be found in \citep{ORTBOBNIK}. A particular feature of the one reported in \citep{ORTBOBNIK} is that it preserves its alertness, that is, it is able to track parameter variations still ensuring the FCT property.\footnote{It should be pointed out that, as is well-known \citep{SLOLIbook,TAObook}, the LS algorithm loses alertness and the covariance matrix needs to be reset to track parameter variations---even under persistent excitation. On the other hand, the HG-FCT does not loose alertness and can track parameter variations.
} All of these schemes involve major variations to the underlying estimation scheme---either gradient search or LS. Our objective with the present note is to bring to the readers attention the fact that it is possible to ensure FCT with the {\em classical} LS scheme, without any such modification.
 
\bibliographystyle{plainnat}
\bibliography{refs}   
\end{document}